# Effect of saccades in tongue electrotactile stimulation for vision substitution applications

A. Chekhchoukh, N. Vuillerme, Y. Payan, N. Glade

*Abstract*— The visual substitution paradigm aims to facilitate the life of blind people. Generally one uses electro-stimulating devices where electrodes are arranged into arrays to stimulate the skin or the tongue mucosa to send signals of visual type to the subjects. When an electro-stimulation signal is applied continuously (*e.g.* when static visual scenes are displayed for a long period of time), the receptors of the affected region can get saturated and the patient may lose the displayed information. We propose here some mechanisms that ameliorate the quality of perception of the electro-stimulation information. The electrical signal is encoded as 2D scenes projected onto the tongue via a Tongue Display Unit, *i.e.* an electro-tactile stimulator formed by a 12x12 matrix of electrodes. We propose to apply stochastic saccades on this signal. Our assumption is that this eye-inspired mechanism should make the visual substitution more efficient (by improving the perception) because of the reduction of the tactile receptors saturation. The influence of saccades was evaluated by a series of experiments. Results revealed a benefit on the persistence of perception due to saccades. This work helps to prevent the saturation of receptors on the tongue, Therefore increasing the quality of vision by the way of the electro-stimulation. It allows new enhancement features to retinal prosthesis devices which suffer from the same phenomenon.

## I. INTRODUCTION

The "visual substitution" paradigm was introduced and extensively studied by Paul Bach-Y-Rita and colleagues in the context of tactile visual substitution systems [1]. These authors evidenced that stimulus characteristics of one sensory modality (e.g., a visual stimulus) could be transformed into stimulations of another sensory modality (e.g., a matrix of vibro-tactile or electro-tactile stimulations on different parts of the body). The "visual substitution" paradigm suggested converting a light energy (a 2D image) into a mechanical or electrical energy (a "tactile image"). Thanks to brain plasticity, this tactile information can be perceived as a visual stimulus. In the recent years, there have been various applications of visual substitution using electro-stimulation on the tongue by the way of low resolution matrices of electrodes. The first tactile visual substitution system (TVSS) was designed by Paul Bach-y-Rita by the end of 1960's. It was composed of a vibro-tactile matrix (20 x 20) placed on the chest or the abdomen of the subject and a video camera that captures the scenes. The rendered images were to be sent to the electro-tactile matrix once the intensity was adjusted and the resolution reduced [1]. The human tongue is one of the most sensitive organs for tactile information with a large cortex mapping which is the reason why Bach-y-Rita and colleagues developed the Tongue Display Unit (TDU) [2]. This device represents a non-invasive brain machine interface. Current version is equipped by a 12 x 12 or 6 x 6 matrix of electrodes that stimulate the dorsal part of the tongue. Yet, numerous published works have confirmed that we can see with the tongue by electro-stimulation, and have proved that the TDU can restore some vision tasks like, shape recognition and object detection. The effectiveness of electrotactile tongue-placed biofeedback systems further has been demonstrated in improving proprioceptive and balance abilities [23, 24] under various (normal or degraded) sensory conditions and neuromuscular states. Other related studies where the TDU was used were successful like guiding surgeons in a specific surgical gesture [17] or reducing abnormal overpressures in seated posture [5, 6]. Using the TDU, object tracking manipulations were done in our experimental platform where the subjects follow a target in front of a stereoscopic camera that recording the movement of the tagged hand [3].

Persons who have lost photoreceptors of their eyes by retina degeneration or age-related macular degeneration (AMD) are susceptible to benefit from prosthetic vision. Implantable retinal prosthetic indeed restores part of the vision by stimulating the retinal neurons or ganglion cells. Several applications were developed using 4 x 4 matrices of lights (phosphenes) as retinal prostheses [11-14]. This technology has become more mature during the last years, with higher resolution (8 x 8), e. g. the Argus II prosthesis produced by Second Sight Company. Furthermore, other simulation projects in laboratories aim to increase resolution and found new image rendering in low resolution using Head Mounting Display [19].

In the listed applications, subjects suffer of major inconvenient in particular the saturation of receptors after short moments of stimulation (electrical or luminous signals) especially in static scenes. In normal vision, mechanisms like micro-saccades, drifts or sparkle help in accomplishing vision tasks properly. Human eyes move permanently to refresh the photo-receptors and resolve the saturation problem that induces a fading effect [20]. Micro-saccades is one of the most important mechanisms for such a role. They indeed contribute (i) to maintain the vision by refreshing the retinal mosaic of photo-receptors, (ii) to help in the perception of low contrasts differences, and (iii) to enhance stereoscopic hyper-acuity [12, 13]. The same mechanism of

A. Chekhchoukh is with the Univ. Grenoble-Alpes, FRE 3405 AGIM Laboratory CNRS/UJF/EPHE/UMPF, Faculty of Medicine of Grenoble, 38700 La Tronche, France. Tel : +33(0)476637153. E-mail: abdes-salem.chekhchoukh@imag.fr

N; Vuillerme is with the Univ. Grenoble-Alpes, FRE 3405 AGIM Laboratory CNRS/UJF/EPHE/UMPF, Faculty of Medicine of Grenoble, 38700 La Tronche, France and Institut Universitaire de France. E-mail: nicolas.vuillerme@agim.eu

Y. Payan is with UJF-Grenoble1/CNRS/TIMC-IMAG UMR5525, Grenoble, F-38041, France. E-mail: yohan.payan@imag.fr

N; Glade is with the Univ. Grenoble-Alpes, FRE 3405 AGIM Laboratory CNRS/UJF/EPHE/UMPF, Faculty of Medicine of Grenoble, 38700 La Tronche. E-mail: nicolas.glade@agim.eu

receptor saturations is probably also observed as concerns the tongue. After a short duration of electro-stimulation on its surface, the corresponding receptors could be saturated, which would alter the individual's perceptive ability to accurately identify and localize the electro-tactile stimulations. The quality of the perception would therefore probably benefit from a refreshing phase of the electro-tactile signals. Unfortunately, such mechanisms that should increases perception were not sufficiently studied by the community working on vision substitution. The geometry of the electrodes has been extensively studied by Kaczmarek et al [8]. Many configurations were proposed and the most comfortable pattern and shape was selected to be used for the current TDU matrices. In the context of prosthetic vision, various models of phosphenes were simulated and a comparative study was reported by Spencer et al [18]. Recently, the threshold of electrical intensity perceived by the tongue receptors was studied by Mitchell et al [15] : these authors defined a spatial mapping of the perceived intensities on the tongue. The concept of "calibration matrix" was proposed to make more uniform the perception in all the stimulated regions on the tongue [3,4]. In a previous study, we showed that eye inherited mechanisms like saccades and sparkle may influence the quality of perception of the orientation of single static lines. In particular the sparkling of the electrotactile tongue stimulations was shown to significantly increase the subjects' sensitivity to the orientation of the stimulations [4].

The aim of the current research is to investigate the effect of the saccades mechanism (inherited from eye micro-saccades) onto the quality of the perception (and therefore probably against saturation phenomena produced by continuous electrical signals). Indeed, we would like to extend the previous study [4] including constructed scenarios with a new series of experiment. The ability of subjects to perceive is measured in the case of different stimulation durations. Series of psychomotor experiments were designed to try to evaluate the threshold of duration beyond which saturation phenomena may occur. We think that this study on visual substitution by electro-stimulation on the tongue will provide some clues to enhance substitution through the retina with new paradigm as concerns the integration of saccades and sparkle in the signals sent to the prosthesis.

## II. MATERIAL & METHODS :

### A. Material :

To specifically conduct our experiments, we have developed an original dedicated experimental platform. This experimental platform was designed to integrate the questions/answers system in form of a scenario. It was made of two main components:

(i) a TDU made of 12 x 12 electrodes matrix (30mmx30mm), the electrodes having a 1.5 mm diameter and being spaced by 2.34 mm center to center. The electro-tactile stimuli scenes are applied to the dorsal part of the tongue [figure 1.a]. A range of 1 to 10 Volts is provided by each electrode, and a calibration matrix is applied to get a comfortable and uniform perception [3, 4].

(ii) A graphics tablet WaCom [fig 1.b] is used to collect the answers in the form of drawing lines points to compute the standard error of angles.

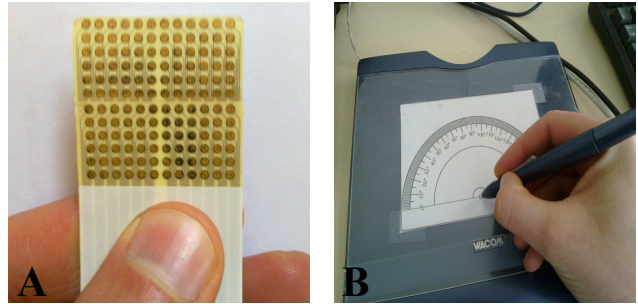

Figure 1. *Devices used in the experiment*: *(A)* Tongue Display Unit composed of 12x12 electrodes. *(B)* Graphics Tablet WaCom, and pen used to draw lines (answers in the form of angles).

### B. Calibration and parameters:

In previous study, the subjects were able to adjust the calibration matrix before each session. In the present study an average matrix was used [4]. This calibration matrix was evaluated and tested to be applied for all the subjects of this study.

### C. Stimuli:

The stimulation starts with a drawing of two crossed lines displayed onto the electro-stimulation matrix (fig.2). This visual scene is sent to the TDU matrix during 2, 5 and 10 seconds. A beep is played when the duration of stimulation is out. After this warning one of the two lines disappears and the subject is asked to draw on the graphical tablet the line that he perceives as remaining on the TDU. The subject's response is saved as the angle of the line drawn on the tablet.

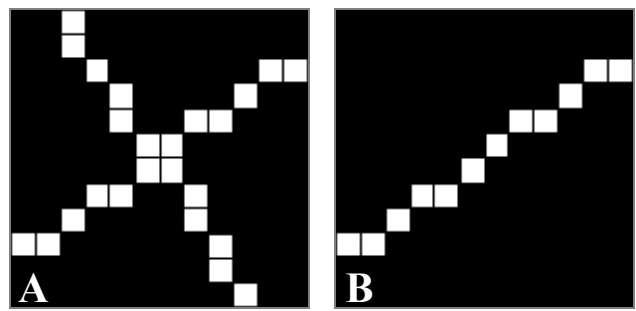

Figure 2. *Stimuli as they appear to the subject*: *(A)* Crossed lines stimulation pattern. After duration T = 2, 5 and 10 seconds, one of the two lines disappears. *(B)* The subject's task was to draw on the graphic tablet the line they perceived through the tongue tactile stimulation.

### D. Subjects :

Seven healthy young male subjects (age 21-33 years old) voluntary participated in this experiment. They were naive to the experience purpose, and used the tongue tactile stimulation for the first time. They were blind folded during the experiment.

*E. Experimental procedure :*

As a start, two crossed lines are displayed with the same intensity during time T. The angle between both lines is fixed to 90°. Two types of tongue electro-tactile stimulation were provided : (1) without saccades (static scene) and (2) with saccades (the taxels that compose the scene move randomly from their original position to its neighborhood, The occurring frequency of saccades is fixed to 0.5 [4]), each of them being displayed during three distinct stimulation durations D : D=2 seconds and D=5 seconds, assuming to correspond to "short" tongue tactile stimulation (<10 seconds), and D=10 seconds, corresponding to "long" tongue tactile stimulation. Longer stimulations were intended to increase the saturation phenomenon of the tongue tactile receptors.

After time D, one of the two lines disappears randomly and a beep signal is played while the chronometer is started to measure the response time. The subject' task is then to draw the perceived line on the graphic tablet as accurately as possible. For each trial, the response data file contains the error between the expected and the answered angles, the response time and the type of stimulation (with or without saccades).

The stimulation angles are comprised between 0° and 360°. We normalized them into angles comprised between 90° and -90° that represent the same orientations. The standard deviations (SD) of angles were computed by subtracting the normalized stimulation angles and the answered angles. We used violin plots to illustrate the distributions of angle standard deviations depending on the presence of saccades and on the duration of stimulation. In particular, we analyzed from these distributions the effect of these two factors on their medians (more representative than means). We used a Mood test to compare the medians of the angle SD distribution.

For each factor (duration of stimulation, presence of saccades), 10 values were collected, for a total of 60 measurements for each subject by session. Two sessions spaced by a period of one to six days were programmed for all the subjects, resulting in 120 measurements in total by subject. As a total, this study is therefore based on 840 measurements performed on 7 subjects.

III. RESULTS

Figure 3 shows the associated violin plots of the SD distributions, obtained in the four different experimental conditions (short or long durations of tongue electrotactile stimulation, with or without saccades).

One can observe notably the increase of standard deviations for long stimulation times without applying saccades, and the positive effect of saccades on the perception.

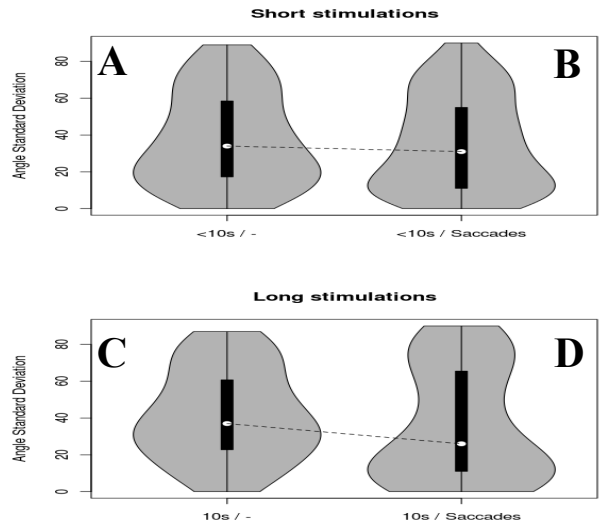

Figure 3. *Distribution of angle standard deviations*: *Short stimulations (<10 seconds) without (A) and with saccades (B). Long stimulations (10 seconds) without (C) and with (D) saccades.*

The decrease on angle SD due to saccades is clearly shown in fig. 3: its value is 26° when saccades are present and first stimulus duration is long (10 seconds), and is equal to 37° when saccades are absent for the same duration of stimulation. A more synthetic view of the results is given in table 1.

TABLE I. *MOOD TEST RESULTS*: SIGNIFICANT P-VALUES ASSOCIATED TO THE COMPARISON OF ANGLE STANDARD DEVIATIONS WERE FOUNDED IN THE PRESENCE OF SACCADES AND TIME STIMULATION OF 10 SECONDS.

| Stimulation time (seconds), and Presence of saccades (yes, no) | Z- score | p-value |
|---|---|---|
| (<10, No) *vs* (<10, Yes) | -0.51 | 0.61 |
| (=10, Yes ) *vs* (<10, Yes) | -2.84 | 0.0046 |
| (<10, No) *vs* (=10, No) | 1.06 | 0.29 |
| (=10, No) *vs* (=10, Yes) | -3.49 | 0.00024 |

When comparing angle SD in the case stimulation is shorter than 10 seconds (2 or 5 seconds) and no saccades are applied with the same duration (<10 seconds) (Fig.3A) and saccades are applied (Fig.3B), no significant change is observed (P>0.05), although the distribution looks slightly better.

In the second condition (10 seconds of stimulation with saccades (Fig.3D) compared to short stimulation without saccades (Fig.3C)), a significant amelioration is observed (smaller median) (P<0.0001). Similar results are observed in the absence of saccades, when comparing short (Fig.3A) and long (Fig.3C) tongue electrotactile stimulation. As seen before in table 1, the p-value associated to the comparison between experiment with (Fig. 3D) and without saccades for long stimulation (Fig.3A) is very significant (*P*<0.0001). This result confirms the positive effect of the presence of saccades in the perception of electric signals for long tongue electrotactile stimulation duration.

## IV. Discussion

The present study investigated the effect of saccades on the improvement of the perception of images during electro-stimulation of the tongue. The long duration of tongue electrotactile stimulation is probably saturating the stimulated regions and the corresponding perception quality considerably decreases. We assume that after 10 seconds of stimulation, the tongue tactile receptors are saturated and subjects have difficulties to determine the orientation of the remaining line. Loomis et al have used slight motion called jitter to prevent the tactile image from fading or losing definition in case of static visuo-tactile scenes [10]. Results obtained in these series of experiments confirm that the saccades participate on refreshing the receptors in such a way the subjects restore part of the perception. By applying saccades to the stimulus, the stimulated receptors are not the same all time since the signal (the lines drawn) is moving around an average position.

One can expect nevertheless that saccades do decrease the resolution of the signal (the signal is drawn all around its average position and each taxel appears then larger) and might consequently affect the quality of perception. This is perhaps the reason why angle SD does not change when applying saccades in conditions of short stimulation. The effect of saccades appears however very significant when the duration of stimulation is long *i.e.* when the receptors of the tongue are presumably more saturated. We think this eye-inspired mechanism is an important feature that should be added to complete the vision substitution paradigm.

Finally, our results should probably be applied now to other sense substitution processes like prosthetic vision where the receptors of the retina might also saturate under persistent electro or luminous stimulation. Saccades should therefore have an effect in prosthetic vision in which eye movements are no longer used since images grabbed by a camera stimulate directly the retina via a matrix of electrodes. In this context, our experimental protocol coupled with a Head Mounting Display could be used to test such processes in prosthetic vision applications.